\documentclass[twocolumn,showpacs,preprintnumbers,amsmath,amssymb]{revtex4}

\usepackage{graphicx}
\usepackage{dcolumn}
\usepackage{bm}

\begin{document}


\title{The stability and free expansion of a dipolar Fermi gas}
\author{L. He$^1$, J.-N. Zhang$^1$, Yunbo Zhang$^2$, and S. Yi$^1$}

\affiliation{$^1$Institute of Theoretical Physics, Chinese Academy
of Sciences, Beijing 100080, China}

\affiliation{$^2$Department of Physics and Institute of Theoretical Physics,
Shanxi University, Taiyuan 030006, China}

\begin{abstract}
We investigate the stability and the free expansion of a trapped dipolar Fermi gas. We show that stabilizing the system relying on tuning the trap geometry is generally inefficient. We further show that the expanded density profile always gets stretched along the attractive direction of dipolar interaction. We also point out that by switching off the dipolar interaction simultaneously with the trapping potential, the deformation of momentum distribution can be directly observed.
\end{abstract}

\date{\today}

\maketitle

{\em Introduction}. --- Since the predication of dipolar effects in a Bose-Einstein condensate~\cite{dbec}, dipolar condensates have drawn significant interests during the past few years~\cite{dip}. The anisotropic nature and the associated tunability of dipole-dipole interaction provide us an unique platform to study the novel quantum phenomena. In particular, shortly after the experimental realization of the dipolar condensate of Cr atoms~\cite{pfau}, experimentalists now have control over both the contact interaction and the trap geometry, and have demonstrated the collapse and stabilization of the Cr condensate~\cite{pfau2}.

The theoretical study on ultra-cold dipolar Fermi gases was initiated by G\'{o}ral {\it et al}.~\cite{goral}. They analyzed the equilibrium state properties and its stability based on a variational approach. The subsequential work includes investigating the BCS pairing originating from the attractive dipolar interaction~\cite{pair} and the strongly correlated states of rotating dipolar Fermi gases~\cite{fqh}. In general, for fermionic atoms, the Pauli exclusion may completely wash out the dipolar effects originating from the weak magnetic dipole-dipole interaction~\cite{goral}. However, with the fast development in making ultra-cold polar molecules~\cite{pmol}, we expect the dipole-dipole interaction to play an important role in determining the fundamental properties of the ultra-cold fermionic polar molecules.

In this paper, we investigate the equilibrium state properties and the free expansion of a polarized dipolar Fermi gas. Based on the full numerical calculations, we show that the critical dipolar interaction strength is significantly lower than that predicted variationally, and it has a very weak dependence on the geometry of the trapping potential. We therefore propose to stabilize a dipolar Fermi gas by tuning the dipolar interaction strength via a rotating orienting field. We then show that, independent of trap geometry, the Fermi gas is always stretched along the attractive direction of dipolar interaction during free expansion. Similar to the dipolar condensates~\cite{pfau,fexp,giov2}, the anisotropic expansion can be used as a diagnostic tool for dipole-dipole interaction.

During the preparation of the manuscript, we become aware of a preprint by Miyakawa {\it et al}. \cite{miya}. They revisited the equilibrium state properties of a dipolar Fermi gas using a variational ansatz capable of describing the deformation in momentum space. It was shown that the exchange dipolar interaction induced the deformation in momentum distribution and lowered the stability of the dipolar Fermi gas. We remark that in the treatment presented in this work, the momentum distribution is assumed to be isotropic, and consequently, the exchange interaction plays no role in determining the equilibrium density and the dynamics. We shall determine the validity region of our work and point out that the deformation in momentum distribution can be measured experimentally by switching off the dipolar interaction during free expansion.

{\em Model}. --- We consider a system of $N$ fermionic polar molecules with permanent dipole moment $d$ at zero temperature. For simplicity, we assume that all dipoles are polarized by an external electric field ${\mathbf E}$ which forms an angle $\varphi$ to $z$-axis. To make the dipolar interaction tunable, we further assume that the orienting electric field fast rotates around $z$-axis. The time average of the dipole-dipole interaction potential becomes~\cite{giov}
\begin{eqnarray}
V_d({\mathbf r})=c_d{\cal V}_{\mathbf 1}({\mathbf r})=c_d\frac{x^2+y^2-2z^2}{(x^2+y^2+z^2)^{3/2}}
\end{eqnarray}
where $c_d=\eta d^2/(4\pi\varepsilon_0)$ with $\eta=(3\cos^2\varphi-1)/2$ being a parameter continuously tunable within the range $[-\frac{1}{2},1]$. The inclusion of $\eta$ not only makes the sign of dipolar interaction changeable, it can also be used to completely switch off the dipolar interaction if $\varphi$ equals to 54.7$^\circ$, the `magic angle'.

\begin{figure}
\centering
\includegraphics[width=2.7in]{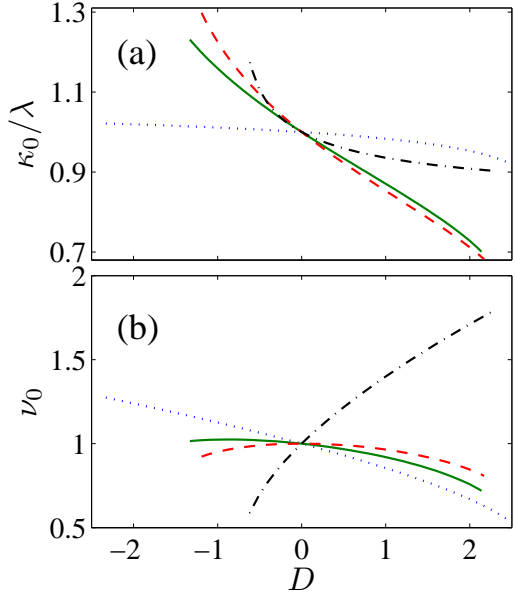}
\caption{The dipolar interaction dependence of the density deformation (a) and the volume of the cloud (b) for $\lambda=0.1$ (dotted line), $0.7$ (solid line), $1$ (dashed line), and $10$ (dash-dotted line).} \label{aratio}
\end{figure}

Within the semi-classical framework, the state of the system is described by the phase space distribution function $f({\mathbf r},{\mathbf v},t)$ which, in the collisionless regime, satisfies the Boltzman-Vlasov kinetic equation~\cite{string}
\begin{eqnarray}
\frac{\partial f}{\partial t}+{\mathbf v}\cdot\frac{\partial f}{\partial{\mathbf r}}-\frac{1}{m}\frac{\partial (U_{\rm ho}+U_{\rm dd})}{\partial{\mathbf r}}\cdot\frac{\partial f}{\partial{\mathbf v}}
=0,\label{bveo}
\end{eqnarray}
where $U_{\rm ho}({\mathbf r})=\frac{1}{2}m\sum_{j=x,y,z}\omega_j^2r_j^2$ is the trapping potential and for simplicity, we assume that $U_{\rm ho}$ is axially symmetric with $\omega_x=\omega_y=\omega_\perp$, $U_{\rm dd}({\mathbf r})=c_d\int d{\mathbf r}'{\cal V}_{\mathbf 1}({\mathbf r}-{\mathbf r}')n({\mathbf r}')$ is the Hartree-Fock mean-field term originating from the dipole-dipole interaction, and $n({\mathbf r})=\int d{\mathbf v}f({\mathbf r},{\mathbf v},t)$ is the density distribution function. Since directly solving Eq. (\ref{bveo}) is generally inapplicable, we shall make use of the scaling ansatz defined as
\begin{eqnarray}
f({\mathbf r},{\mathbf v},t)=f_0({\mathbf R}(t),{\mathbf V}(t)),
\end{eqnarray}
where $R_j(t)=r_j/b_j(t)$, $V_j(t)=b_j(t)v_j-\dot b_j(t)r_j$, and $f_0$ is the equilibrium phase space distribution function. This scaling transformation was first introduced to study the free expansion of a Bose-Einstein condensate~\cite{castin}, and the collective oscillations of a classical Bose gas~\cite{guery}. Recently, it was also generalized to study the free expansion and the collective excitation of both normal and superfluid Fermi gases~\cite{string,hu}. Following the standard procedure~\cite{guery,string,hu}, the Boltzman-Vlasov equation (\ref{bveo}) can be reduced to the coupled equations for the scaling parameters $b_j$, which, in dimensionless form~\cite{dimless}, reads
\begin{eqnarray}
\ddot b_j+\lambda_j^2b_j-\frac{\lambda_j^2}{ b_j^3}-\frac{D{\cal T}_{j}({\mathbf 1})}{\langle R_j^2\rangle b_j^3}+\frac{D{\cal T}_{j}({\mathbf b})}{\langle R_j^2\rangle b_j}=0,\label{dyn}
\end{eqnarray}
where $\lambda_j=\omega_j/\omega_\perp$, $D=\eta N^{1/6} \sqrt{m^3\omega_\perp/\hbar^5}\frac{d^2}{4\pi\varepsilon_0}$
is a dimensionless quantity characterizing the strength of dipolar interaction, and $\langle R_j^2\rangle=\int d{\mathbf R}R_j^2n_0({\mathbf R})$ with $n_0({\mathbf R})=\int d{\mathbf V}f_0({\mathbf R},{\mathbf V})$ being the equilibrium density. For the axially symmetric trap studied in this work, we have $\langle R_x^2\rangle=\langle R_y^2\rangle$. Furthermore,
\begin{eqnarray}
{\cal T}_{j}({\mathbf b})=\int d{\mathbf R}d{\mathbf R}'R_jn_0({\mathbf R}){\cal V}_{\mathbf b}({\mathbf R}-{\mathbf R}')\frac{\partial n_0({\mathbf R}')}{\partial R_j'}
\end{eqnarray}
where
\begin{eqnarray}
{\cal V}_{\mathbf b}({\mathbf R})=\frac{ b_x^2X^2+ b_y^2Y^2-2 b_z^2Z^2}{( b_x^2X^2+ b_y^2Y^2+ b_z^2Z^2)^{5/2}}\nonumber
\end{eqnarray}
is the dipole-dipole interaction potential under the scaling transformation. Once the equilibrium density $n_0$ is obtained, the dynamics of the system can be studied by evolving Eqs. (\ref{dyn}). Before presenting our results on free expansion, we would like to address the equilibrium state properties of a trapped dipolar Fermi gas.

{\em Equilibrium density}. --- Under local density approximation, the equilibrium density is determined by equation \cite{goral}
\begin{eqnarray}
\mu&=&\frac{1}{2}\left(6\pi^2n_0\right)^{2/3}+\frac{1}{2}(x^2+y^2+\lambda^2z^2)\nonumber\\
&&+D\int d{\mathbf r}'{\cal V}_{\mathbf 1}({\mathbf r}-{\mathbf r}')n_0({\mathbf r}'),\label{equi}
\end{eqnarray}
where $\lambda\equiv\lambda_z$ and the chemical potential $\mu$ is introduced to ensure that $n_0$ is normalized to unit. The properties of equilibrium density was previously analyzed by G\'{o}ral {\it et al}.~\cite{goral} using variational approaches. Here we would like to tackle the problem numerically. Utilizing the cylindrical symmetry of the system and the numerical technique developed by Ronen {\it et al}.~\cite{ronen}, we are able to obtain $n_0$ with very high precision. The numerical solutions are further checked using Virial theorem~\cite{goral}.

\begin{figure}
\centering
\includegraphics[width=2.7in]{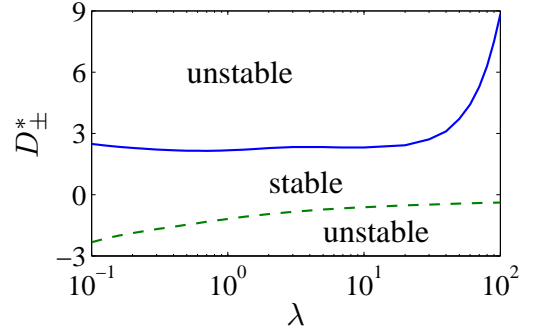}
\caption{The $\lambda$ dependence of the critical dipolar interaction strength $D_+^*>0$ (solid line) and $D_-^*<0$ (dashed line).} \label{cri}
\end{figure}

To characterize the properties of the equilibrium density, we define the aspect ratio of the density profile as $\kappa_0=\sqrt{\langle R_x^2\rangle/\langle R_z^2\rangle}$. Since $\kappa_0=\lambda$ for a noninteracting Fermi gas,  $\kappa_0/\lambda$ measures the density deformation induced by dipole-dipole interaction. Figure \ref{aratio} (a) shows the dipolar interaction dependence of $\kappa_0/\lambda$. We immediately see that the density profile always gets stretched along the attractive direction of the dipolar interaction, i.e., the axial (radial) direction for positive (negative) $D$. In general, this result does not hold true for a dipolar condensate, except that the condensate falls into the Thomas-Fermi regime due to the large repulsive contact interaction~\cite{yi4}. As the dipolar interaction is always partially attractive, the system becomes unstable if the dipolar interaction exceeds certain critical strength. In other words, a dipolar Fermi gas can only be stable when $D_-^*<D<D_+^*$. To understand how the system collapses, we introduce the reduced `volume' of the cloud $\nu_0=\frac{\langle R_x^2\rangle\langle R_z^2\rangle^{1/2}}{\lambda^{-1/2}(3/32)^{1/2}}$, which becomes unit in the absence of the dipolar interaction. In Fig. \ref{aratio} (b), we plot the $D$ dependence of $\nu_0$ for various trap geometries. Generally, $\nu_0$ increases (decreases) with $|D|$ if the overall dipolar interaction is repulsive (attractive). For near spherical trap, $\nu_0$ always decreases with $|D|$ when the dipolar interaction approaches the critical values. However for highly elongate or oblate traps, the system becomes unstable even when the overall interaction is repulsive. This observation suggests that in these cases the collapse is initiated locally, which is also confirmed in our numerical simulation.

\begin{figure}
\centering
\includegraphics[width=2.7in]{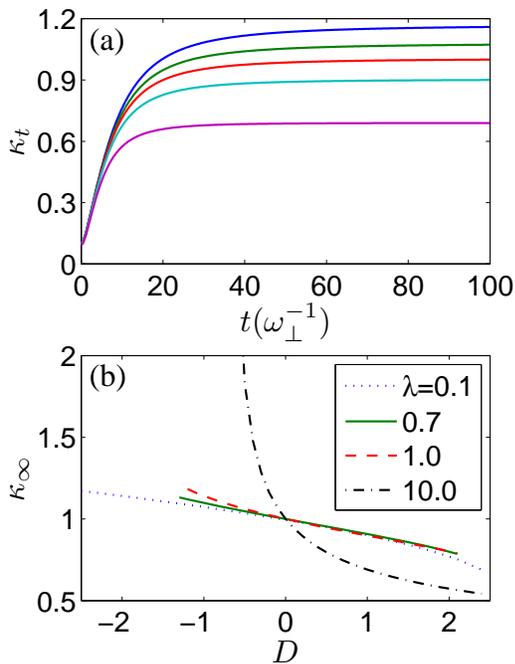}
\caption{(a) The time dependence of cloud aspect ratio for $\lambda=0.1$. In descending order, the dipolar interaction parameters are $D=-2.3$, $-1$, $0$, $1$, and $2.4$. (b)The dipolar interaction dependence of asymptotical aspect ratios for various $\lambda$'s.} \label{free}
\end{figure}

In Fig. \ref{cri}, we present the $\lambda$ dependence of the critical dipolar interaction strength. $|D_-^*|$ monotonically decreases as one increases $\lambda$, indicating that a more pronounced cigar-shaped trap stabilizes the dipolar Fermi gas with negative $D$. This can be easily understood as the dipolar interaction is repulsive along $z$-axis. On the other hand, $D_+^*$ barely depends on the trap geometry for $\lambda<20$, and it only shows significant increase as we further increase $\lambda$. Compared to the variational results~\cite{goral}, the stability gained by increasing $\lambda$ is very limited and $D_+^*$ always remains finite. The large discrepancy between the variational and the numerical predications on stability boundaries stems from the fact that a simple variational ansatz is generally incapable of capturing the local collapse. Similar discrepancy between the variational and numerical results also exists for dipolar Bose-Einstein condensates~\cite{yi4,yi3,ronen2}.

Based on above discussion, stabilizing a dipolar Fermi gas by increasing $\lambda$ is inefficient. Consider, for example, a typical polar molecule with $d=1\,{\rm  Debye}$ and $m=100\,{\rm a.m.u}$, and the radial trap frequency $\omega_\perp=(2\pi)100$ Hz, one reaches $D\approx1.5\eta N^{1/6}$. Therefore, without tuning $\eta$, a highly oblate trap with $\lambda=100$ can only sustain about $N\simeq 4.7\times 10^4$ fermions. Taking into account the deformation of momentum distribution may further lower the critical number of molecules~\cite{miya}. More efficiently, we may tune $|\eta|$ to a small value to stabilize the system. Alternatively, we may also load the polar molecules into a 1D optical lattice, which not only provides an highly oblate trap, but also divides the system into many subsystems containing fewer molecules.

{\em Free expansion}. --- To study the free expansion, we numerically evolve Eqs. (\ref{dyn}) with the restoring force term, $\lambda_j^2b_j$, being removed. In Fig. \ref{free} (a), we present the time dependence of the cloud aspect ratio $\kappa_t=\kappa_0 b_x(t)/b_z(t)$. For a noninteracting Fermi gas, $\kappa_t$ always reaches unit at the end of expansion, independent of the trap geometry. In the presence of dipolar interaction, $\kappa_t$ quickly reaches its asymptotical value $\kappa_\infty$ which deviates from 1. Figure \ref{free} (b) shows the $D$ dependence of the asymptotical aspect ratio. One immediately sees that independent of the trap geometry and the sign of $D$, the expanded cloud always gets stretched along the attractive direction of dipolar interaction. At first sight, this result may look counterintuitive, but it can be understood as the dipole-dipole mean-field potential accelerates the particles along attractive dipolar interaction direction, while decelerates those along repulsive direction~\cite{giov2}. It is also worth mentioning that, for polar molecules, the deformation in the expanded density should be readily observed in experiment.

\begin{figure}
\centering
\includegraphics[width=2.7in]{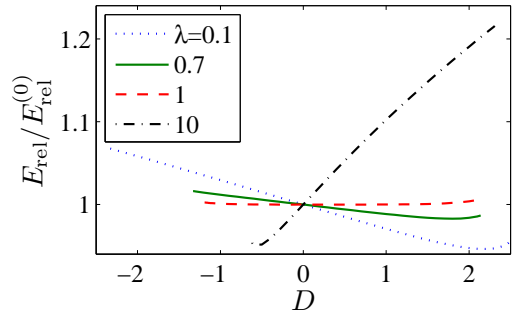}
\caption{The $D$ dependence of the release energy for various $\lambda$'s.} \label{erel}
\end{figure}

In addition to the expanded density profile, the release energy $E_{\rm rel}$, sum of the kinetic and interaction energies after switching off the trap, can also be measured experimentally. The release energy in the absence of dipolar interaction is $E_{\rm rel}^{(0)}=\frac{3}{8}(6\lambda)^{1/3}$ for a Fermi gas. The typical behavior of the release energy is presented in Fig. \ref{erel}. For a highly oblate trap, the release energy may differ significantly from that of a noninteracting Fermi gas.

\begin{figure}
\centering
\includegraphics[width=2.7in]{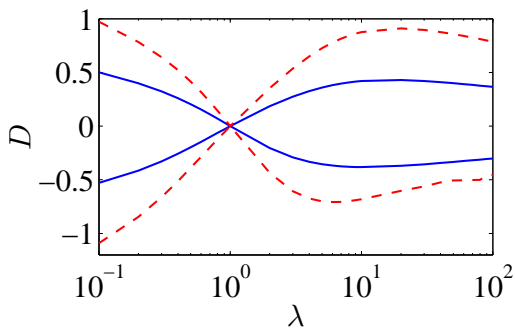}
\caption{Plot of contour lines on $\lambda$-$D$ parameter plane for $E_{\rm ex}/E_{\rm dd}=0.05$ (solid line) and $0.1$ (dashed line). The direct and exchange dipolar interactions ratio for $(\lambda, D)$ enclosed by the contour line is lower than that on the corresponding contour line.} \label{eex}
\end{figure}

{\em Deformation of momentum distribution}. --- As we have mentioned, the results presented above are obtained based on the assumption that the momentum distribution is isotropic. Consequently, the exchange dipole-dipole interaction vanishes in our treatment. In Ref. \cite{miya}, Miyakawa {\it et al}. however showed that, due to the anisotropy of dipolar interaction, the exchange interaction may induce the deformation in momentum space, such that the momentum distribution is always stretched along the attractive direction of the dipolar interaction. It is therefore important for us to compare the direct and exchange dipolar interactions in order to validate our calculations. In Fig. \ref{eex}, we presented the direct and exchange dipolar interaction ratio $E_{\rm dd}/E_{\rm ex}$ based on the variational calculation developed in Ref.~\cite{miya}. It can be seen that the exchange interaction has most dramatic effect when the trapping potential is near spherical, and it becomes less important in highly prolate and oblate traps. The reason behind this is that, compared to the density distribution, the momentum distribution only weakly depends on the trap geometry.

The deformation in momentum distribution also gives rise to an interesting question about how to observe it directly. Since interaction always contaminates momentum distribution during expansion, we have to turn it off during expansion. This can actually be achieved by tuning $\eta$ to zero simultaneously with the switch-off of the external trapping potential, such that the gas expands ballistically.

{\em Conclusion}. --- We have studied the equilibrium state properties and the free expansion of a dipolar Fermi gas based on the semi-classical theory. From the stability diagram, we propose to stabilize a gas of fermionic polar molecules by directly tuning the dipolar interaction via a fast rotating field. We have also shown that the expanded gas always get stretched along the attractive direction of the dipolar interaction. For a typical polar molecular formed by alkali atoms, the deformation in expanded density profile can be readily detected. Finally, we point out that the deformation of momentum distribution can be directly observed by switching off the dipolar interaction during free expansion. We hope that our work will stimulate experimental efforts along this line. Our future work will include the study of the collective excitations of trapped dipolar Fermi gases.

We thank Han Pu, Duan-lu Zhou, and Ruquan Wang for the helpful discussion. This work is supported by NSFC (Grant No. 10774095), a fund from National 973 program (Grant No. 2006CB921102), and the ``Bairen" program of Chinese Academy of Sciences.


\begin{thebibliography}{}
\bibitem{dimless} Here we have adopted following units: $a_\perp=N^{1/6}\sqrt{\hbar/m\omega_\perp}$ for length, $\omega_\perp^{-1}$ for time, $N^{4/3}\hbar\omega_\perp$ for energy, and $N/a_\perp^3$ for density.
\end{thebibliography}
\end{document}